\documentclass[published]{JHEP3} 
\JHEP{00(2010)000}

\JHEPspecialurl{http://jhep.sissa.it/JOURNAL/JHEP3.tar.gz}

\usepackage{epsfig,multicol,bbm}

\newcommand\fverb{\setbox\fverbbox=\hbox\bgroup\verb}
\newcommand\fverbdo{\egroup\medskip\noindent%
			\fbox{\unhbox\fverbbox}\ }
\newcommand\fverbit{\egroup\item[\fbox{\unhbox\fverbbox}]}
\newbox\fverbbox

\title{Could $Y_{b}(10890)$ be the P-wave $[bq][\bar{b}\bar{q}]$ tetraquark state?}

\author{Jian-Rong Zhang and Ming-Qiu Huang\\

Department of Physics, National University of Defense
Technology, Hunan 410073, China\\
E-mail: \email{jrzhang@nudt.edu.cn, mqhuang@nudt.edu.cn}}

\received{} 		
\accepted{}		

\abstract{Assuming $Y_{b}(10890)$
as a P-wave $bq$-scalar-diquark $\bar{b}\bar{q}$-scalar-antidiquark tetraquark state,
the mass of $Y_{b}(10890)$ is
computed in the framework of QCD
sum rule method. Technically, contributions of operators up to
dimension six are included in the operator product expansion (OPE). The
numerical result $10.88\pm0.13~\mbox{GeV}$ for
$Y_{b}(10890)$ agrees well with the experimental value, which favors the P-wave
$[bq][\bar{b}\bar{q}]$ tetraquark configuration for $Y_{b}(10890)$.
In the same picture, the mass of $Y(4360)$ is calculated and the result $4.32\pm0.20~\mbox{GeV}$
is compatible with the experimental data, which supports $Y(4360)$'s P-wave
$[cq][\bar{c}\bar{q}]$ structure.}


\begin{document}
\section{Introduction}\label{sec1}
The observations of $\Upsilon(1S)\pi^{+}\pi^{-}$ and $\Upsilon(2S)\pi^{+}\pi^{-}$
states near the $\Upsilon(5S)$ resonance \cite{Y10890,Y10890-new} have attracted
great theoretical attention \cite{Y10890-theory}.
However, there are still
some puzzles on the anomalously large rates and the way to describe distribution shapes
and the helicity angle.
Recently, Ali {\it et al.} \cite{tetraquark}
identify $Y_{b}(10890)$ with the state $Y_{[bq]}(10900)$ \cite{Y10900}
and interpret $Y_{b}(10890)$ as a P-wave $[bq][\bar{b}\bar{q}]$ tetraquark
state.
In this way, a dynamical model for decays
$Y_{b}\rightarrow\Upsilon(1S)\pi^{+}\pi^{-}$,
$\Upsilon(2S)\pi^{+}\pi^{-}$ is presented, which provides
excellent fits for the decay distributions.
Therefore, it is interesting to investigate whether $Y_{b}(10890)$
could be a tetraquark state.
Undoubtedly,
the quantitative description of $Y_{b}(10890)$'s properties like mass is
helpful for understanding its structure, but it is
difficult to extract the hadronic spectrum information from the
simple QCD Lagrangian.
That is because low energy QCD
involves a regime where it is futile to attempt perturbative
calculations and one has to treat a genuinely strong field in
nonperturbative methods.
However, one can apply QCD sum rules \cite{svzsum} (for reviews see
\cite{overview,overview1,overview2,overview3} and references
therein), which
are a nonperturbative formulation firmly rooted in QCD.
From the above reasons, we devote to study
$Y_{b}(10890)$ with QCD sum
rules in this work.

Additionally, BABAR Collaboration observed a broad structure $Y(4325)$ in
the process $e^{+}e^{-}\rightarrow\gamma_{ISR}\pi^{+}\pi^{-}\psi(2s)$
at $4324\pm24~\mbox{MeV}$ with a width $172\pm33~\mbox{MeV}$ \cite{Y4325}.
Latterly,
Belle Collaboration reported the charmoniumlike
state $Y(4360)$ in $e^{+}e^{-}\rightarrow\pi^{+}\pi^{-}\psi(2s)$
at $4361\pm9\pm9~\mbox{MeV}$ with a width of $74\pm15\pm10~\mbox{MeV}$
\cite{Y4360}.
The mass of $Y(4325)$ is close to that of $Y(4360)$, and the main
difference between them is their widths.
It seems very difficult to observe these two structures
simultaneously because of the large width of $Y(4325)$.
They could be the same structure and the width difference may be
due to the experimental error. In Ref. \cite{Y4360-fit}, Liu {\it et al.} have tried to perform a combined fit to
$e^{+}e^{-}\rightarrow\pi^{+}\pi^{-}\psi(2s)$ cross sections measured by the
BABAR and Belle experiments. In this
work, we assume $Y(4325)$ and $Y(4360)$ are exactly the
same resonance for simplicity. On $Y(4360)$,
there have already been some theoretical works \cite{Y4360-theory,Y4360-theory1,Y4360-theory2}. From QCD sum rules,
Albuquerque {\it et al.} arrive at
$M=4.49\pm0.11~\mbox{GeV}$, adopting the current
$[cq]_{S=0}[\bar{c}\bar{q}]_{S=1}+
[cq]_{S=1}[\bar{c}\bar{q}]_{S=0}$ \cite{Y4360-QCDSR} (for the concise review
on multiquark QCD sum rules, one can see
\cite{overview-MN}).
At present,
we would like to
study whether $Y(4360)$ could be a P-wave $[cq][\bar{c}\bar{q}]$ tetraquark state.

The paper is organized as follows. In Sec. \ref{sec2}, the QCD sum rule for the tetraquark state
is introduced, and both the
phenomenological representation and QCD side are derived, followed
by the numerical analysis to extract the hadronic masses in Sec.
\ref{sec3}. Section \ref{sec4} is a brief summary.

\section{The tetraquark state QCD sum rule}\label{sec2}
In the tetraquark interpretation, $Y_{[Qq]}$ is a $J^{PC}=1^{--}$ diquark-antidiquark state,
having the flavor content $[Qq][\bar{Q}\bar{q}]$. Its spin and orbital momentum
numbers are: $S_{[Qq]}=0$, $S_{[\bar{Q}\bar{q}]}=0$, $S_{[Qq][\bar{Q}\bar{q}]}=0$, and $L_{[Qq][\bar{Q}\bar{q}]}=1$ \cite{diquark}.
For the interpolating current, a derivative could be included to generate $L_{[Qq][\bar{Q}\bar{q}]}=1$.
Presently, one constructs the tetraquark state current
from diquark-antidiquark
configuration of fields, while constructs the molecular state current from
meson-meson type of fields.
Although these two types of currents can be related to each other by Fiertz rearrangements,
the relations are suppressed
by color and Dirac factors \cite{overview-MN}.
It will have a maximum
overlap for the tetraqurk state using the
diquark-antidiquark current and
the sum rule can reproduce the physical mass well.
Thus, the following form of current could be
constructed for the P-wave $[Qq][\bar{Q}\bar{q}]$,
\begin{eqnarray}
j^{\mu}=\epsilon_{abc}\epsilon_{dec}(q_{a}^{T}C\gamma_{5}Q_{b})D^{\mu}(\bar{q}_{d}\gamma_{5}C\bar{Q}_{e}^{T}).
\end{eqnarray}
Here the index $T$ means matrix
transposition, $C$ is the charge conjugation matrix, $D^{\mu}$ denotes the covariant derivative, and $a$, $b$,
$c$, $d$, and $e$ are color indices.

The mass sum rule starts from the two-point correlator
\begin{eqnarray}\label{correlator}
\Pi^{\mu\nu}(q^{2})=i\int
d^{4}x\mbox{e}^{iq.x}\langle0|T[j^{\mu}(x)j^{\nu\dagger}(0)]|0\rangle.
\end{eqnarray}
Lorentz covariance implies that the correlator (\ref{correlator})
can be generally parameterized as
\begin{eqnarray}
\Pi^{\mu\nu}(q^{2})=\bigg(\frac{q^{\mu}q^{\nu}}{q^{2}}-g^{\mu\nu}\bigg)\Pi^{(1)}(q^{2})+\frac{q^{\mu}q^{\nu}}{q^{2}}\Pi^{(0)}(q^{2}).
\end{eqnarray}
The part proportional to $g_{\mu\nu}$ is
chosen to extract the sum rule here. In phenomenology,
$\Pi^{(1)}(q^{2})$ can be expressed as a dispersion integral
\begin{eqnarray}
\Pi^{(1)}(q^{2})=\frac{[\lambda^{(1)}]^{2}}{M_{H}^{2}-q^{2}}+\frac{1}{\pi}\int_{s_{0}}
^{\infty}ds\frac{\mbox{Im}\Pi^{(1)\mbox{phen}}(s)}{s-q^{2}}+\mbox{subtractions},
\end{eqnarray}
where $M_{H}$ denotes the mass of the hadronic resonance. In the OPE
side, $\Pi^{(1)}(q^{2})$ can be written in terms of a dispersion
relation as
\begin{eqnarray}
\Pi^{(1)}(q^{2})=\int_{4m_{Q}^{2}}^{\infty}ds\frac{\rho^{\mbox{OPE}}(s)}{s-q^{2}},
\end{eqnarray}
where the spectral density is given by
\begin{eqnarray}
\rho^{\mbox{OPE}}(s)=\frac{1}{\pi}\mbox{Im}\Pi^{\mbox{(1)}}(s).
\end{eqnarray}
After equating the two sides, assuming quark-hadron duality, and
making a Borel transform, the sum rule can be written as
\begin{eqnarray}\label{sumrule}
[\lambda^{(1)}]^{2}e^{-M_{H}^{2}/M^{2}}&=&\int_{4m_{Q}^{2}}^{s_{0}}ds\rho^{\mbox{OPE}}(s)e^{-s/M^{2}}.
\end{eqnarray}
To eliminate the hadronic coupling constant $\lambda^{(1)}$,
one reckons the ratio of derivative of the sum rule and itself, and then
yields
\begin{eqnarray}\label{sum rule}
M_{H}^{2}&=&\int_{4m_{Q}^{2}}^{s_{0}}ds\rho^{\mbox{OPE}}s
e^{-s/M^{2}}\bigg/
\int_{4m_{Q}^{2}}^{s_{0}}ds\rho^{\mbox{OPE}}e^{-s/M^{2}}.
\end{eqnarray}

To calculate the OPE side, we
work at leading order in $\alpha_{s}$ and
consider condensates up to dimension six with the
same techniques in Refs. \cite{technique,technique1}.
To keep the heavy-quark
mass finite, one uses the momentum-space expression
for the heavy-quark propagator, including two and three gluons
attached expressions given in Ref. \cite{reinders}.
The light-quark part of the correlator
is calculated in the
coordinate space and then Fourier-transformed to the momentum
space in $D$ dimension.
The resulting light-quark part
is combined with the heavy-quark part and
dimensionally regularized.
It is defined as $r(m_{Q},s)=(\alpha+\beta)m_{Q}^2-\alpha\beta s$ and $K(\alpha,\beta)=1+\alpha-2\alpha^{2}+\beta+2\alpha\beta-2\beta^{2}$.
The spectral density is written as
\begin{eqnarray}
\rho^{\mbox{OPE}}(s)&=&\rho^{\mbox{pert}}(s)+\rho^{\langle\bar{q}q\rangle}(s)+\rho^{\langle\bar{q}q\rangle^{2}}(s)+\rho^{\langle
g\bar{q}\sigma\cdot G q\rangle}(s)+\rho^{\langle
g^{2}G^{2}\rangle}(s)+\rho^{\langle g^{3}G^{3}\rangle}(s),\nonumber\\
\rho^{\mbox{pert}}(s)&=&-\frac{1}{3\cdot5\cdot2^{11}\pi^{6}}\int_{\alpha_{min}}^{\alpha_{max}}\frac{d\alpha}{\alpha^{4}}\int_{\beta_{min}}^{1-\alpha}\frac{d\beta}{\beta^{4}}(1-\alpha-\beta)K(\alpha,\beta)
r(m_{Q},s)^{5},\nonumber\\
\rho^{\langle\bar{q}q\rangle}(s)&=&\frac{m_{Q}\langle\bar{q}q\rangle}{3\cdot2^{6}\pi^{4}}\int_{\alpha_{min}}^{\alpha_{max}}\frac{d\alpha}{\alpha^{2}}\int_{\beta_{min}}^{1-\alpha}\frac{d\beta}{\beta^{2}}(2-\alpha-\beta)r(m_{Q},s)^{3},\nonumber\\
\rho^{\langle\bar{q}q\rangle^{2}}(s)&=&-\frac{m_{Q}^{2}\langle\bar{q}q\rangle^{2}}{3\cdot2^{3}\pi^{2}}\int_{\alpha_{min}}^{\alpha_{max}}d\alpha\Big[m_{Q}^{2}-\alpha(1-\alpha)s\Big],\nonumber\\
\rho^{\langle g\bar{q}\sigma\cdot G q\rangle}(s)&=&-\frac{m_{Q}\langle
g\bar{q}\sigma\cdot G
q\rangle}{2^{8}\pi^{4}}\int_{\alpha_{min}}^{\alpha_{max}}\frac{d\alpha}{\alpha^{2}}\int_{\beta_{min}}^{1-\alpha}\frac{d\beta}{\beta^{2}}(\alpha+\beta-4\alpha\beta)r(m_{Q},s)^{2}\nonumber\\
& &{}
+\frac{m_{Q}\langle
g\bar{q}\sigma\cdot G
q\rangle}{2^{8}\pi^{4}}\int_{\alpha_{min}}^{\alpha_{max}}\frac{d\alpha}{\alpha(1-\alpha)}\Big[m_{Q}^{2}-\alpha(1-\alpha)s\Big]^{2},\nonumber\\
\rho^{\langle g^{2}G^{2}\rangle}(s)&=&-\frac{m_{Q}^{2}\langle
g^{2}G^{2}\rangle}{3^{2}\cdot2^{12}\pi^{6}}\int_{\alpha_{min}}^{\alpha_{max}}\frac{d\alpha}{\alpha^{4}}\int_{\beta_{min}}^{1-\alpha}\frac{d\beta}{\beta^{4}}(1-\alpha-\beta)(\alpha^{3}+\beta^{3})K(\alpha,\beta)
r(m_{Q},s)^{2},~\mbox{and}\nonumber\\
\rho^{\langle g^{3}G^{3}\rangle}(s)&=&-\frac{\langle
g^{3}G^{3}\rangle}{3^{2}\cdot2^{14}\pi^{6}}\int_{\alpha_{min}}^{\alpha_{max}}\frac{d\alpha}{\alpha^{4}}\int_{\beta_{min}}^{1-\alpha}\frac{d\beta}{\beta^{4}}(1-\alpha-\beta)K(\alpha,\beta)
\Big[(\alpha^{3}+\beta^{3})r(m_{Q},s)\nonumber\\&&{}+4(\alpha^{4}
+\beta^{4})m_{Q}^{2}\Big]
r(m_{Q},s).\nonumber
\end{eqnarray}
The integration limits are given
by $\alpha_{min}=(1-\sqrt{1-4m_{Q}^{2}/s})/2$,
$\alpha_{max}=(1+\sqrt{1-4m_{Q}^{2}/s})/2$, and $\beta_{min}=\alpha
m_{Q}^{2}/(s\alpha-m_{Q}^{2})$.
Note that the next-to-leading order corrections are not
included here, for which one needs to consider the renormalization of the current \cite{jin}.
This procedure is undoubtedly complicated and tedious,
since the renormalization of the current will raise
the operator-mixing problems. Especially for the multiquark system,
many operators will mix under renormalization.
Actually,
a lot of hard calculations already need to be done
even if one works at leading order
for the difficulties of tackling with the massive propagator diagrams.
Under such a circumstance, it is expected that one could
obtain a trusty sum rule
working at leading order in $\alpha_{s}$,
and it has been tested to
be feasible for many multiquark states \cite{overview-MN}.
To improve on the accuracy of the QCD sum rule analysis,
it is certainly meaningful to consider the next-to-leading order corrections,
which may be
included in some further work after fulfilling a burdensome task.

\section{Numerical analysis}\label{sec3}

In this Section, the sum rule (\ref{sum rule})
will be numerically analyzed. The input values are taken as
$m_{c}=(1.23\pm0.05)~\mbox{GeV}$, $m_{b}=(4.20\pm0.07)~\mbox{GeV}$, \cite{PDG}
$\langle\bar{q}q\rangle=-(0.23\pm0.03)^{3}~\mbox{GeV}^{3}$, $\langle
g\bar{q}\sigma\cdot G q\rangle=m_{0}^{2}~\langle\bar{q}q\rangle$,
$m_{0}^{2}=0.8~\mbox{GeV}^{2}$, $\langle
g^{2}G^{2}\rangle=0.88~\mbox{GeV}^{4}$, and $\langle
g^{3}G^{3}\rangle=0.045~\mbox{GeV}^{6}$ \cite{overview2,Y4360-QCDSR,overview-MN}.
Complying with the standard criterion
of sum rule analysis, the threshold $s_{0}$ and Borel
parameter $M^{2}$ are varied to find the optimal stability window.
In the QCD sum rule approach,
there is approximation in the OPE of the correlation function, and there is a very
complicated and largely unknown structure of the hadronic dispersion integral in the phenomenological
side.
Therefore, the match of the two sides is not independent
of $M^{2}$.  One expects that there exists a range of $M^{2}$,
in which the two sides have a good
overlap and the sum rule can work well. In practice, one can analyse the convergence in the OPE side and the pole
contribution dominance in the phenomenological side to determine the allowed Borel
window: on one hand, the lower
constraint for $M^{2}$ is obtained by the consideration that the
perturbative contribution should be larger than
the condensate contributions,
to keep the convergence of the OPE under control and insure that one does
not introduce a large error neglecting higher dimension terms; on the other hand,
the upper limit
for $M^{2}$ is obtained by the
restriction that the pole contribution
should be larger
than the QCD continuum contribution, to guarantee that the contributions from high
resonance states and continuum states remains a small part in the phenomenological side.
Meanwhile,
the threshold parameter
$\sqrt{s_{0}}$ is not completely
arbitrary but characterizes the beginning
of the continuum state.
On all accounts,
it is expected that
the two sides have a good
overlap in the determined work window
and information on the resonance can be safely
extracted.

Concretely, the comparison
between the pole and continuum contributions from sum rule (\ref{sumrule}) for $Y_{b}(10890)$
for $\sqrt{s_{0}}=11.5~\mbox{GeV}$ is shown in FIG. 1, and its OPE convergence by comparing different
contributions is shown in FIG. 2. In detail, the
perturbative contribution versus the total OPE contribution at $M^{2}=9.5~\mbox{GeV}^{2}$ is nearly $63\%$, and the ratio increases
with $M^{2}$ to insure that the perturbative contribution can dominate in the total OPE contribution
when $M^{2}\geq9.5~\mbox{GeV}^{2}$. On the other side, at $M^{2}=10.5~\mbox{GeV}^{2}$ the relative pole contribution
is approximately $51\%$, which descends along with $M^{2}$ to guarantee the pole contribution can dominate in the total contribution while $M^{2}\leq10.5~\mbox{GeV}^{2}$. Thus, the regions of $s_{0}$ and $M^{2}$ for $Y_{b}(10890)$ are
taken as $\sqrt{s_0}=11.4\sim11.6~\mbox{GeV}$ and
$M^{2}=9.5\sim10.5~\mbox{GeV}^{2}$. For $Y(4360)$,
the comparison
between the pole and continuum contributions from sum rule (\ref{sumrule})
is shown in FIG. 3, and its OPE convergence by comparing different
contributions is shown in FIG. 4. From the similar analyzing processes,
the regions of $s_{0}$ and $M^{2}$ are
taken as $\sqrt{s_0}=4.8\sim5.0~\mbox{GeV}$ and
$M^{2}=2.6\sim3.6~\mbox{GeV}^{2}$ for $Y(4360)$.
The corresponding Borel curves to determine masses of $Y_{b}(10890)$ and $Y(4360)$ from sum rule (\ref{sum rule})
are shown in FIG. 5 and in FIG. 6, respectively.
Finally, we obtain $10.88\pm0.13~\mbox{GeV}$ for
$Y_{b}(10890)$ and $4.32\pm0.20~\mbox{GeV}$ for $Y(4360)$.
For $Y(4360)$, our central value is closer to
the experimental data comparing with the prediction $4.49\pm0.11~\mbox{GeV}$
in Ref. \cite{Y4360-QCDSR}, however, the uncertainty of our result is larger.

\DOUBLEFIGURE{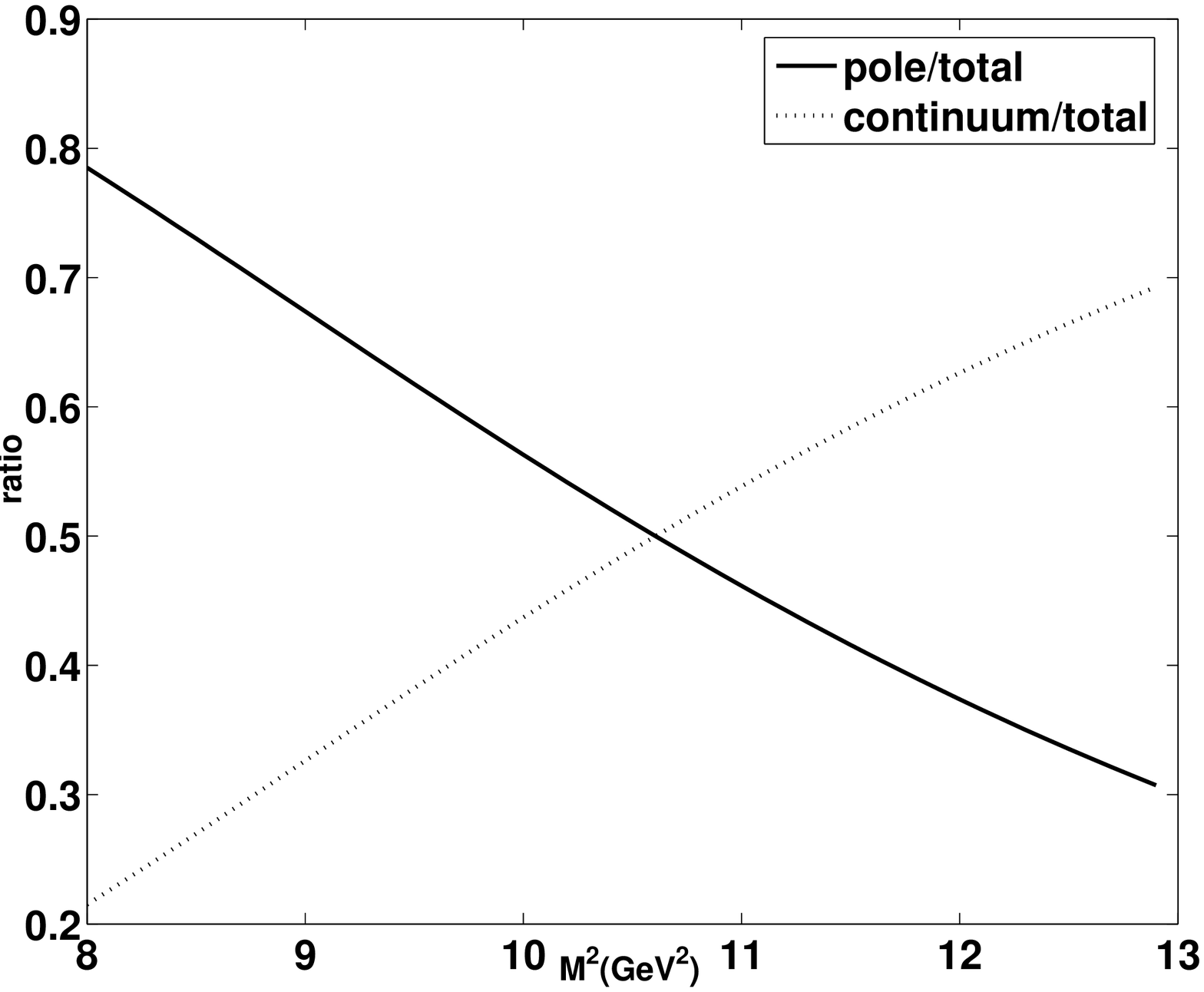, width=.50\textwidth}
{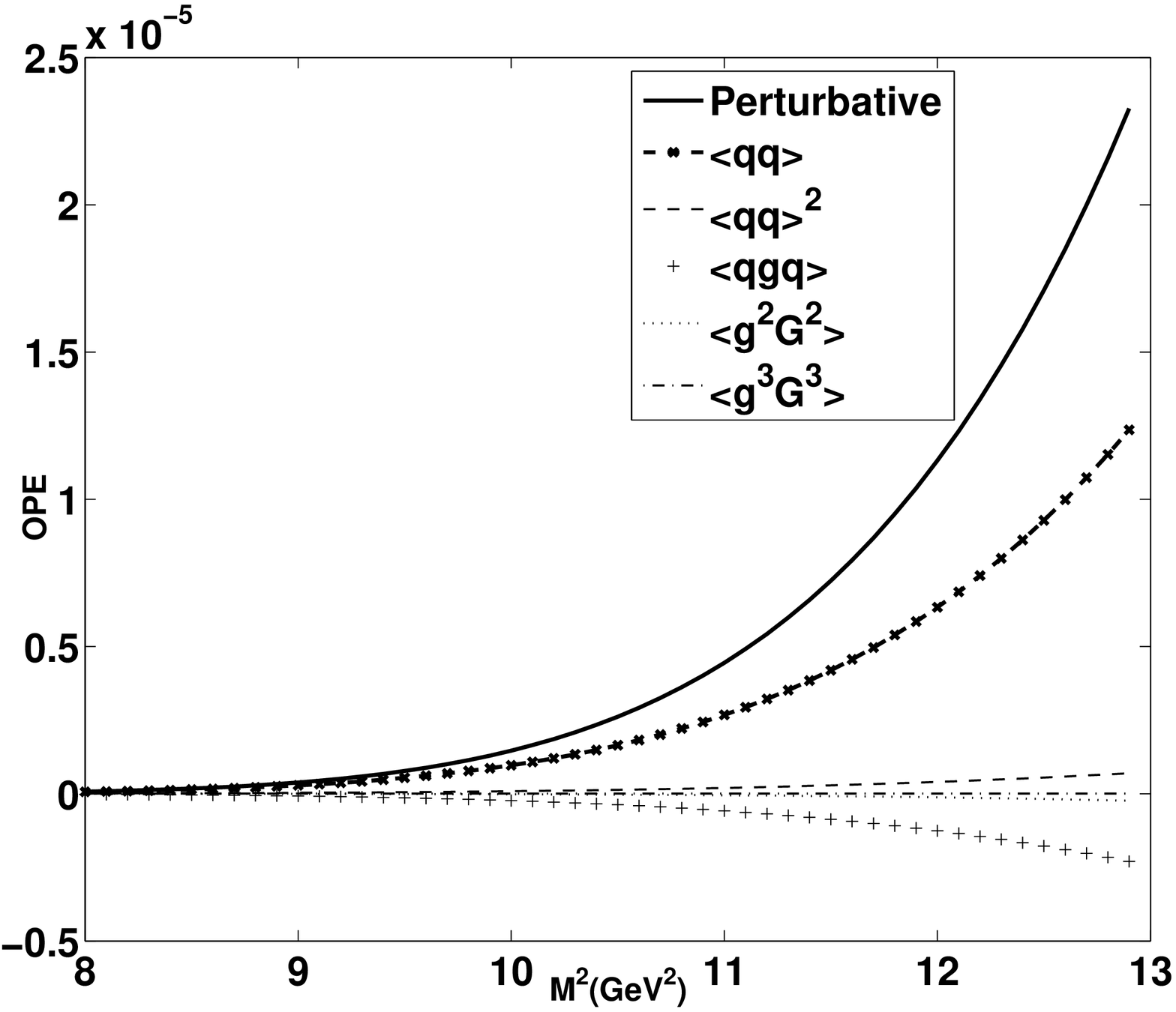, width=.50\textwidth}{The solid line shows the relative pole contribution
(the pole contribution divided by the total, pole plus continuum
contribution) and the dashed line shows the relative continuum
contribution for
$Y_{b}(10890)$.}{The OPE convergence is shown by comparing the
perturbative, quark condensate, four-quark condensate, mixed condensate, two-gluon condensate, and three-gluon
condensate contributions for
$Y_{b}(10890)$.}

\DOUBLEFIGURE{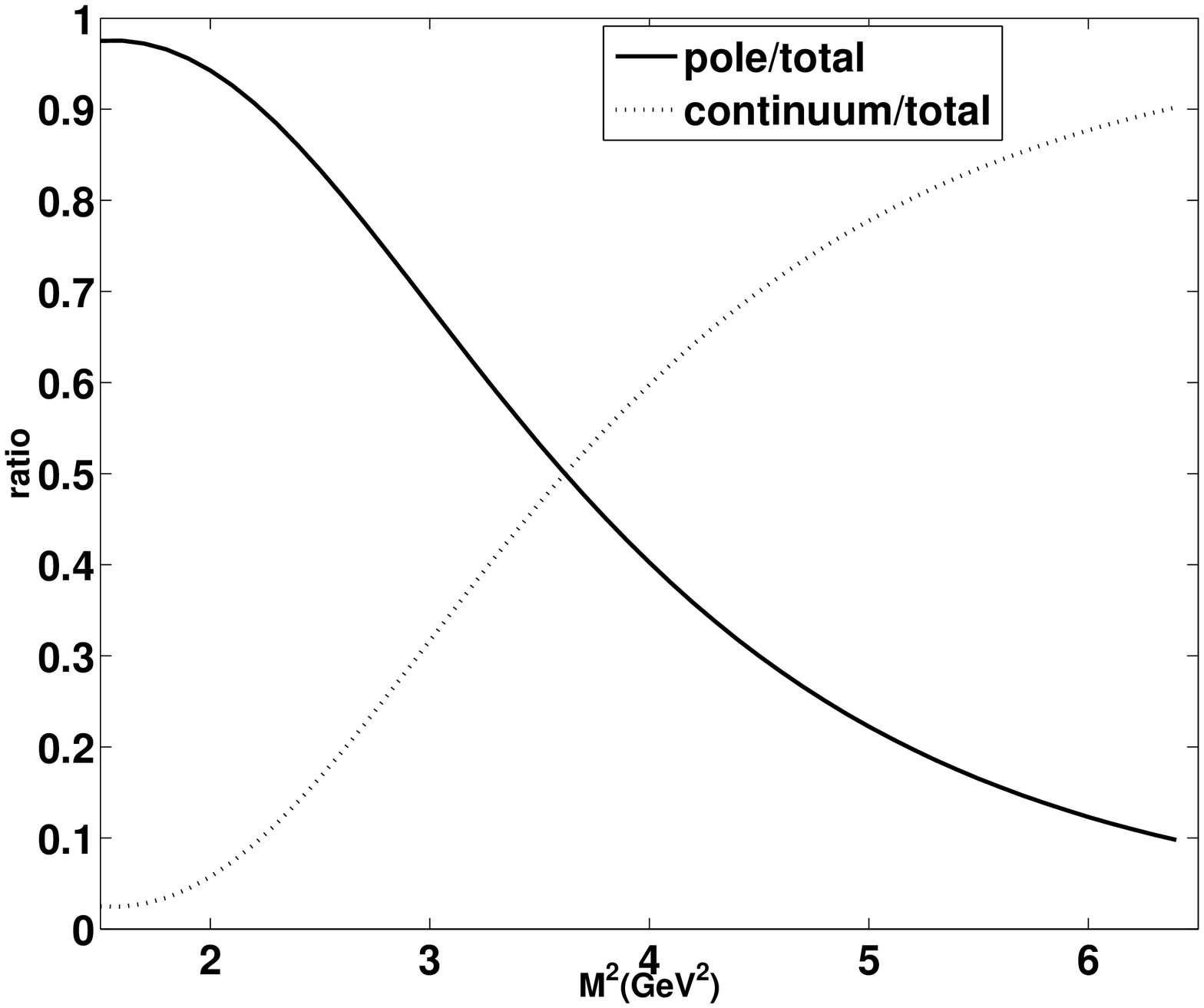, width=.50\textwidth}
{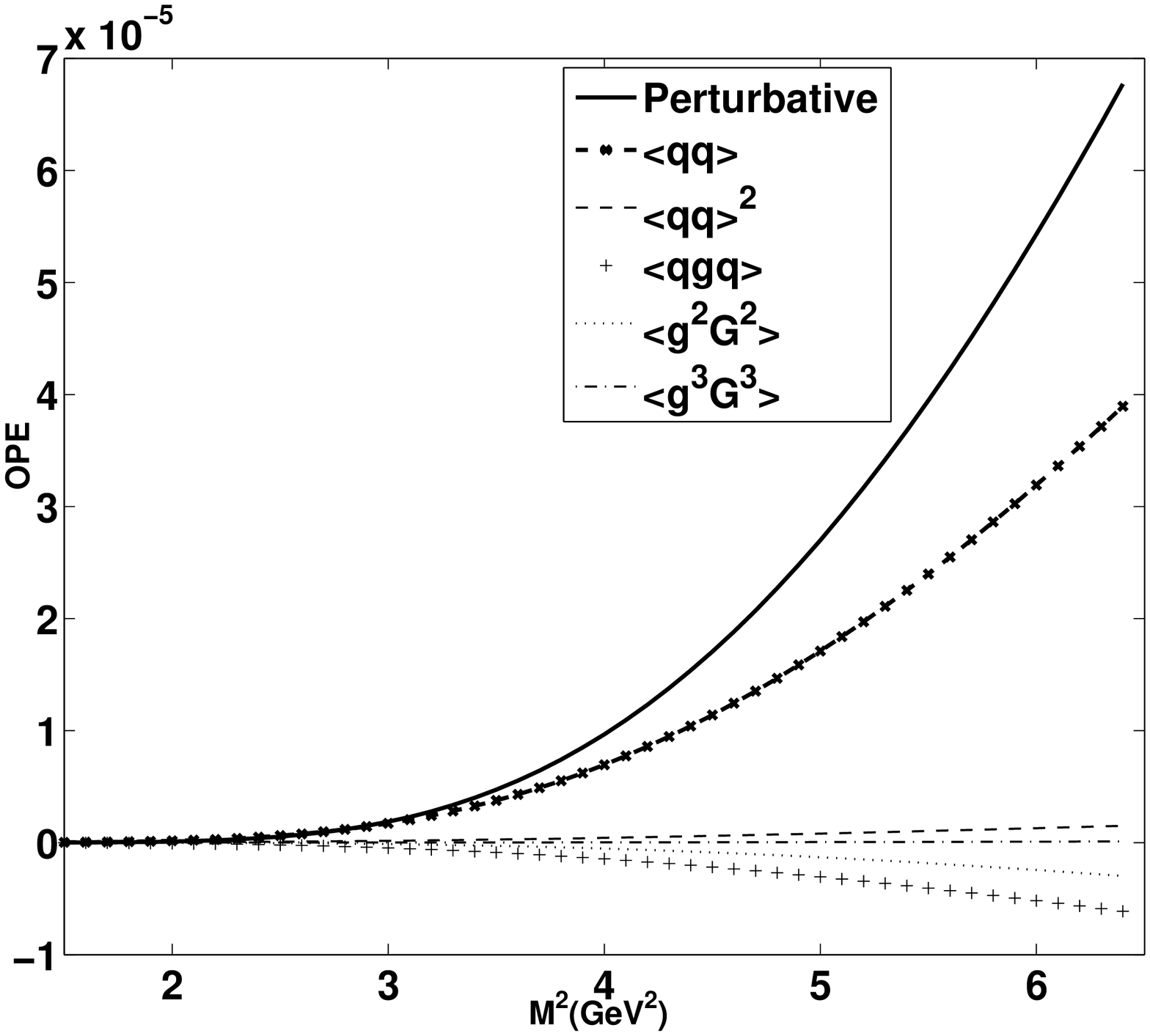, width=.50\textwidth}{The solid line shows the relative pole contribution
(the pole contribution divided by the total, pole plus continuum
contribution) and the dashed line shows the relative continuum
contribution for
$Y(4360)$.}{The OPE convergence is shown by comparing the
perturbative, quark condensate, four-quark condensate, mixed condensate, two-gluon condensate, and three-gluon
condensate contributions for
$Y(4360)$.}

\DOUBLEFIGURE{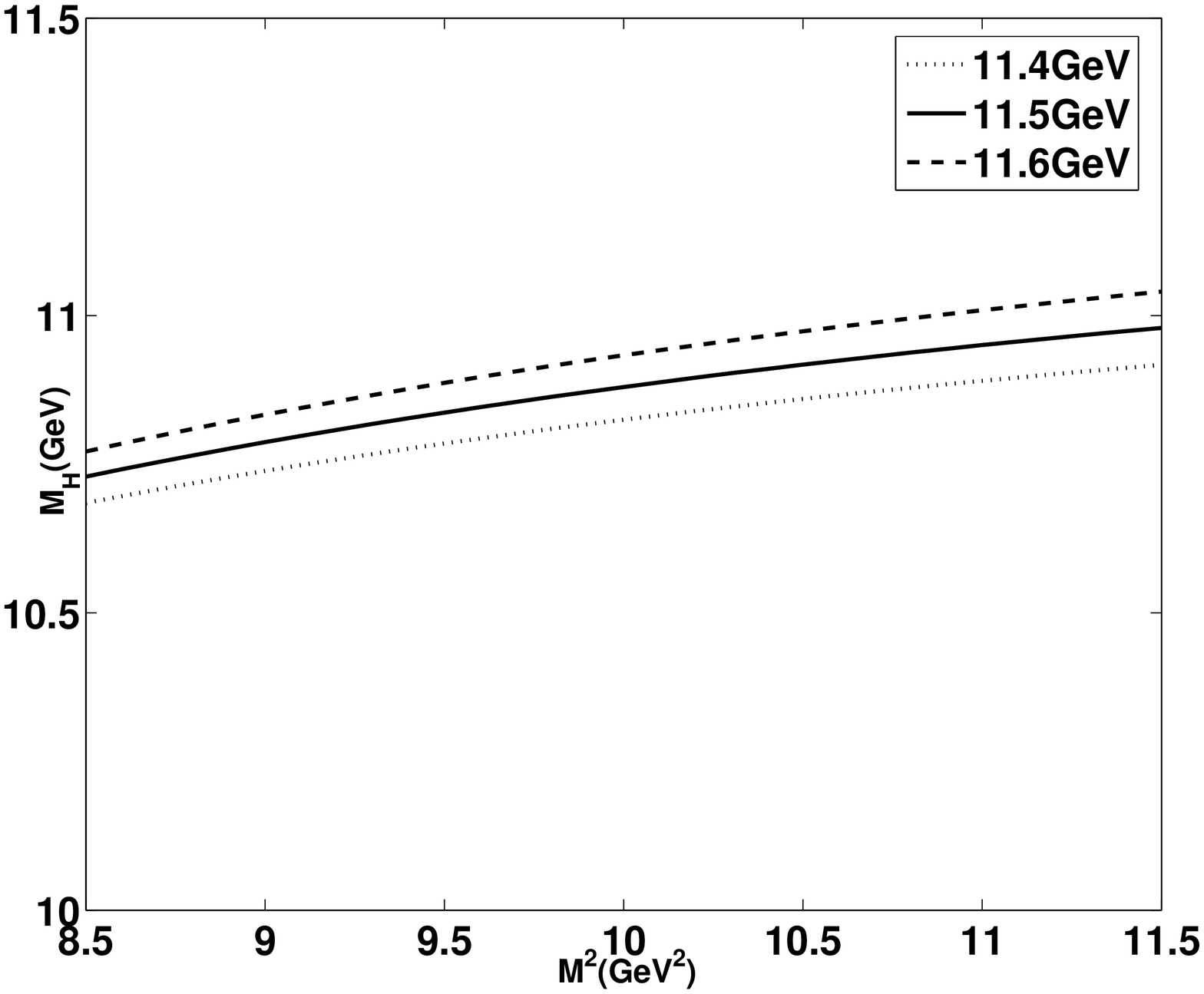, width=.50\textwidth}
{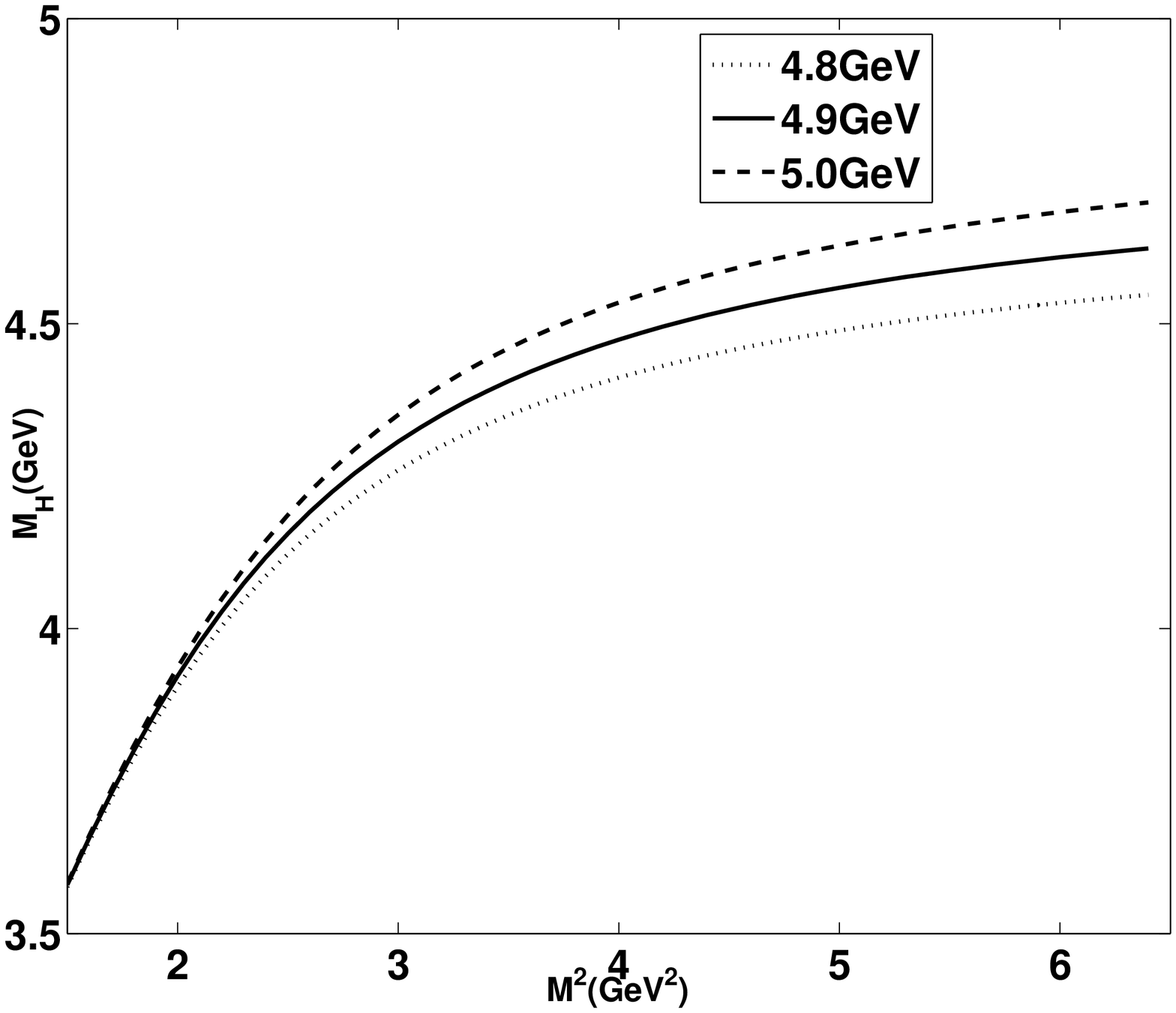, width=.50\textwidth}{The
dependence on $M^2$ for the mass of $Y_{b}(10890)$ is shown. }{The
dependence on $M^2$ for the mass of $Y(4360)$ is shown.}

\section{Summary}\label{sec4}
In the tentative P-wave $[bq][\bar{b}\bar{q}]$ configuration, the QCD sum rule method has been employed to compute the
mass of $Y_{b}(10890)$, including the
contributions of operators up to dimension six in the OPE.
The numerical result $10.88\pm0.13~\mbox{GeV}$ for
$Y_{b}(10890)$ is well compatible with the experimental data, which favors the P-wave tetraquark configuration for $Y_{b}(10890)$.
In the same picture, the mass of $Y(4360)$ has been calculated to be $4.32\pm0.20~\mbox{GeV}$,
and the result is in agreement with the experimental value, which supports its P-wave
$[cq][\bar{c}\bar{q}]$ configuration. We expect the results could be
helpful to understand the structures of these states.
For further work,
one needs to take into
account other dynamical analysis to identify the structures of hadrons.

\begin{acknowledgments}
This work was supported in part by the National Natural Science
Foundation of China under Contract No.10975184.
\end{acknowledgments}

\end{document}